\documentclass[twocolumn,showpacs,amsmath,amssymb,aps,superscriptaddress,nofootinbib,longbibliography]{revtex4-2}
\usepackage[utf8]{inputenc}
\usepackage{graphicx}
\usepackage{physics}
\usepackage{bm}
\usepackage[bookmarks, colorlinks=true, urlcolor=blue, linkcolor=black, citecolor=blue]{hyperref}

\usepackage{todonotes} % for comments

\usepackage[linesnumbered,titlenumbered,ruled,vlined,resetcount]{algorithm2e}
\def\HiLi{\leavevmode\rlap{\hbox to \hsize{\color{green!50}\leaders\hrule height .8\baselineskip depth .5ex\hfill}}}
\SetCommentSty{mycommfont}

\SetKwInput{KwInput}{Input}                
\SetKwInput{KwOutput}{Output}

\usepackage{tcolorbox}
\usepackage{braket}
\usepackage{amsmath}

\usepackage{amssymb}

\usepackage{graphicx}
\usepackage{bbold} 
\usepackage{tikz}
\usepackage{booktabs}

\usepackage{xspace}
\usepackage{xcolor}

\usepackage{appendix}
\usepackage{dsfont}
\usepackage{graphicx}               % Necessary to use \scalebox
\usepackage{mwe}

\usepackage{etoolbox}

\renewcommand{\epsilon}{\varepsilon}

\newtheorem{theorem}{Theorem}

\newtheorem{defn}{Definition} % definition numbers are dependent on theorem numbers
\newcommand{\RomanNumeralCaps}[1]
    {\MakeUppercase{\romannumeral #1}}

\DeclareMathOperator*{\argmax}{arg\,max}

\usepackage[autostyle=true]{csquotes} % Required to generate language-dependent quotes in the bibliography

\begin{document}

% \crefname{appendix}{appendix}{appendices}

\title{Reinforcement Learning Assisted Recursive QAOA}

\begin{abstract}
In recent years, variational quantum algorithms such as the Quantum Approximation Optimization Algorithm (QAOA) have gained popularity as they provide the hope of using NISQ devices to tackle hard combinatorial optimization problems. 
It is, however, known that at low depth, certain locality constraints of QAOA limit its performance. 
To go beyond these limitations, a non-local variant of QAOA, namely recursive QAOA (RQAOA), was proposed to improve the quality of approximate solutions. 
The RQAOA has been studied comparatively less than QAOA, and it is less understood, for instance, for what family of instances it may fail to provide high-quality solutions.
However, as we are tackling $\mathsf{NP}$-hard problems (specifically, the Ising spin model), it is expected that RQAOA does fail, raising the question of designing even better quantum algorithms for combinatorial optimization.
In this spirit, we identify and analyze cases where (depth-1) RQAOA fails and, based on this, propose a reinforcement learning enhanced RQAOA variant (RL-RQAOA) that improves upon RQAOA. 
We show that the performance of RL-RQAOA improves over RQAOA: RL-RQAOA is strictly better on these identified instances where RQAOA underperforms and is similarly performing on instances where RQAOA is near-optimal.
Our work exemplifies the potentially beneficial synergy between reinforcement learning and quantum (inspired) optimization in the design of new, even better heuristics for complex problems.

\end{abstract}

\author{Yash J. Patel}
\thanks{These two authors contributed equally.}
\affiliation{LIACS, Universiteit Leiden, 2333 CA Leiden, The Netherlands}

\author{Sofiene Jerbi}
\thanks{These two authors contributed equally.}
\affiliation{Institute for Theoretical Physics, University of Innsbruck, Technikerstr. 21a, A-6020 Innsbruck, Austria}

\author{Thomas B\"{a}ck}
\affiliation{LIACS, Universiteit Leiden, 2333 CA Leiden, The Netherlands}

\author{Vedran Dunjko}
\affiliation{LIACS, Universiteit Leiden, 2333 CA Leiden, The Netherlands}

\date{\today}
\maketitle

\section{Introduction}

As quantum computing is becoming practical~\cite{google2020hartree, jurcevic2020demonstration, ebadi2021quantum, gong2021quantum}, there has been a growing interest in employing near-term quantum algorithms to help solve problems in quantum chemistry~\cite{moll2018quantum}, quantum machine learning~\cite{benedetti2019parameterized}, and combinatorial optimization~\cite{farhi2014quantum}. 
Any such near-term algorithm must consider the primary restrictions of Noisy Intermediate Scalable Quantum (NISQ) devices; e.g., the number of qubits, decoherence etc.
Variational Quantum Algorithms (VQAs) such as the Quantum Approximation Optimization Algorithm (QAOA)~\cite{farhi2014quantum} were developed as a potential approach to achieve a quantum advantage in practical applications keeping in mind these design restrictions.

For a user-specified input depth \(l\), QAOA consists of a quantum circuit with \(2l\) variational parameters. In the limit of infinite depth, for optimal parameters, the solution of QAOA converges to the optimum for a given combinatorial optimization problem~\cite{farhi2014quantum}.
However, a significant body of research has produced negative results~\cite{hastings2019classical,marwaha2021local, barak2021classical, bravyi2020obstacles, farhi2020mis, farhi2020mc, chou2021limitations, marwaha2021bounds} for QAOA limited to logarithmic depth (in the number of qubits), exploiting the notion of \emph{locality} or \emph{symmetry} in QAOA.
This motivates the study of techniques that circumvent the restriction of locality or symmetry in QAOA, which exploit the information-processing capabilities of low-depth quantum circuits by employing classical non-local pre-and post-processing steps\footnote{The time complexity of these auxiliary steps should be polynomial in input size for the algorithm to remain practically viable.}.

One such proposal is the recursive QAOA (RQAOA), a non-local variant of QAOA, which uses shallow depth circuits of QAOA iteratively, and at every iteration, the size of the problem (usually expressed in terms of a graph or a hypergraph) is reduced by one (or more).
The elimination procedure introduces non-local effects via the new connections between previously unconnected nodes, which counteracts the locality restrictions of QAOA.
The authors in~\cite{bravyi2020obstacles, bravyi2020hybrid, bravyi2021classical} empirically show that depth-1 RQAOA always performs better than depth-1 QAOA and is competitive to best known classical algorithms based on rounding of a semidefinite programming relaxation for Ising and graph colouring problems. 
However, given that these problems are \(\mathsf{NP}\)-hard, there must also exist instances that RQAOA fails to solve exactly, unless \(\mathsf{NP} \subseteq \mathsf{\mathsf{BQP}}\).
Hence, to further push the boundaries of algorithms for combinatorial optimization on NISQ devices (and beyond), it is helpful to determine when RQAOA fails, as this can aid in developing better variants of RQAOA.

In this work, we study extensions of RQAOA, which perform better than RQAOA for the Ising problem (or equivalently, the weighted Max-Cut problem, where the external field is zero, refer to Sec.~\ref{subsec:classical_sim}).
We do this by identifying cases where RQAOA fails (i.e., find small-scale instances with approximation ratio \(\leq 0.95\)).
Then, we analyze the reasons for this failure and, based on these insights, we modify RQAOA.
We employ reinforcement learning (RL) to not only tweak RQAOA's selection rule, but also train the parameters of QAOA instead of using energy-optimal ones in a new algorithm that we call RL-RQAOA.
In particular, the proposed hybrid algorithm provides a suitable test-bed for assessing the potential benefit of RL: we perform simulations of (depth-1) RQAOA, and RL-RQAOA on an ensemble of randomly generated weighted $d$-regular graphs and show that RL-RQAOA consistently outperforms its counterparts. 
In the proposed algorithm, the RL component itself plays an integral role in finding the solution, so this raises the question of the actual role of the QAOA circuit and thus potential quantum advantages.
To show that the QAOA circuits have a non-trivial contribution to the advantage, we compare RL-RQAOA to an entirely classical RL agent (which, given exponential time, imitates a brute force algorithm) and show that RL-RQAOA converges both faster and to better solutions than the simple classical RL agents.
We note that our approach to enhance RQAOA's performance is not limited to depth-1 and can be straightforwardly extended to higher depths.

We present our results as follows: Sec.~\ref{sec:background} introduces QAOA, recursive QAOA (RQAOA), and fundamental concepts behind policy gradient methods in RL.
Sec.~\ref{sec:related_work} presents related works.
Sec.~\ref{sec:rqaoa_lim} describes the limitations of RQAOA, and we illustrate their validity by performing numerical simulations.
In Sec.~\ref{sec:rl_rqaoa}, we provide a sketch of the policies of RL-RQAOA (quantum-classical) and RL-RONE (classical, introduced to characterize the role of quantum aspects of the algorithm) and their learning algorithms. 
Sec.~\ref{sec:numerics} presents our computational results for the comparison between classical and hybrid algorithms (RQAOA, RL-RQAOA, and RL-RONE) on an ensemble of Ising instances.
Finally, we conclude with a discussion in Sec. \ref{sec:disc}.

\section{Background}\label{sec:background}

In this section, we first provide a brief overview of QAOA (Sec.~\ref{subsec:qaoa}) and its classical simulatability for the Ising problem (Sec.~\ref{subsec:classical_sim}). Later, we introduce recursive QAOA (RQAOA) (Sec.~\ref{subsec:rqaoa}) upon which we base our proposal for RL-enhanced RQAOA and introductory concepts behind policy gradient in RL (Sec.~\ref{subsec:rl_primer}). 
These notions will give us tools to develop policies based on the QAOA ansatz and their learning algorithms in the upcoming sections.

\subsection{Quantum Approximate Optimization Algorithm}\label{subsec:qaoa}

QAOA seeks to approximate the maximum of the binary cost function $\mathcal{C}: \{0,1\}^n \rightarrow \mathbb{R}$ encoded into a Hamiltonian as $H_n = \sum_{x \in \{0,1\}^n} \mathcal{C}(x) \ket{x}\bra{x}$. 
Starting from an initial state $\ket{s} = \ket{+^n}$ (uniform superposition state), QAOA alternates between two unitary evolution operators $U_p(\gamma) = \exp(-i \gamma H_n)$ (phase operator) and $U_m(\alpha) = \exp(-i \alpha H_b)$ (mixer operator) respectively, where $H_b = \sum_{j=1}^{n} X_j$. 
Hereafter, $X, Y, Z$ are standard Pauli operators and $P_j$ is a Pauli operator acting on qubit $j$ for $P \in \{X, Y, Z\}$. The phase and mixer operator are typically applied a total of $l$ times, generating the quantum state,
\begin{eqnarray}
        \ket{\Psi_l(\vec{\alpha}, \vec{\gamma})} &=& \prod_{k=1}^l \exp(-i \alpha_k H_b) \exp(-i \gamma_k H_n) \ket{s},
    \label{eq:variationalstate}
\end{eqnarray}
where the variational parameters \{\(\vec{\alpha}, \vec{\gamma}\} \in [0,2\pi]^{2l}\) and the integer $l$ is called the QAOA \emph{depth}. The depth $l$ controls the non-locality of the QAOA circuit. 
During the operation of QAOA, these parameters are tuned to optimize the expected value of $H_n := \bra{\Psi_l(\vec{\alpha}, \vec{\gamma})} H_n \ket{\Psi_l(\vec{\alpha}, \vec{\gamma})}$. 
The preparation of the state (\ref{eq:variationalstate}) is followed by a measurement in the computational basis, which outputs a bitstring \(x\) corresponding to a candidate solution of the cost function \(\mathcal{C}\). 
The probability \(\mathbb{P}_l(x)\) of obtaining a bitstring \(x \in \{0,1\}^n\) is given by Born's rule,
\begin{eqnarray}\label{eqn:prob_bit}
  \mathbb{P}_l(x) =  \left| \langle x|\Psi_l(\vec{\alpha}, \vec{\gamma})\rangle \right|^2.
    \label{eq:prob_bitstring}
\end{eqnarray}

A candidate bitstring \(x^*\) is called an \(r\)-approximation solution to a given instance, for \(0 \leq r \leq 1\) if,
\begin{eqnarray}
    \mathcal{C}(x^*) \geq r \cdot \max_x \mathcal{C}(x).
    \label{eq:apprx}
\end{eqnarray}
An algorithm is said to achieve an approximation ratio of \(r\) for a cost function \(\mathcal{C}\) if it returns an \(r\)-approximation or better for every problem instance in the class (i.e., in the worst case).

We say that depth-$l$ QAOA achieves an approximation ratio of $r$ for a problem instance of a cost function $\mathcal{C}$ if there exists parameters $\{\vec{\alpha}, \vec{\gamma}\}$ such that
\begin{eqnarray}
    \langle H_n \rangle_l := \bra{\Psi_l(\vec{\alpha}, \vec{\gamma})} H_n \ket{\Psi_l(\vec{\alpha}, \vec{\gamma})} \geq r \cdot \max_x \mathcal{C}(x)
    \label{eq:approx_ratio}
\end{eqnarray}

We note that repeating a sequence of state preparations and measurements approximates the distribution of \(x\) given by (\ref{eqn:prob_bit}) and that (\ref{eq:approx_ratio}) is the mean of this distribution. 
The candidate bitstring \(x^*\) may then be selected to yield the maximum approximation ratio \(r\).

\begin{figure*}
    \centering
    \includegraphics[width=2\columnwidth]{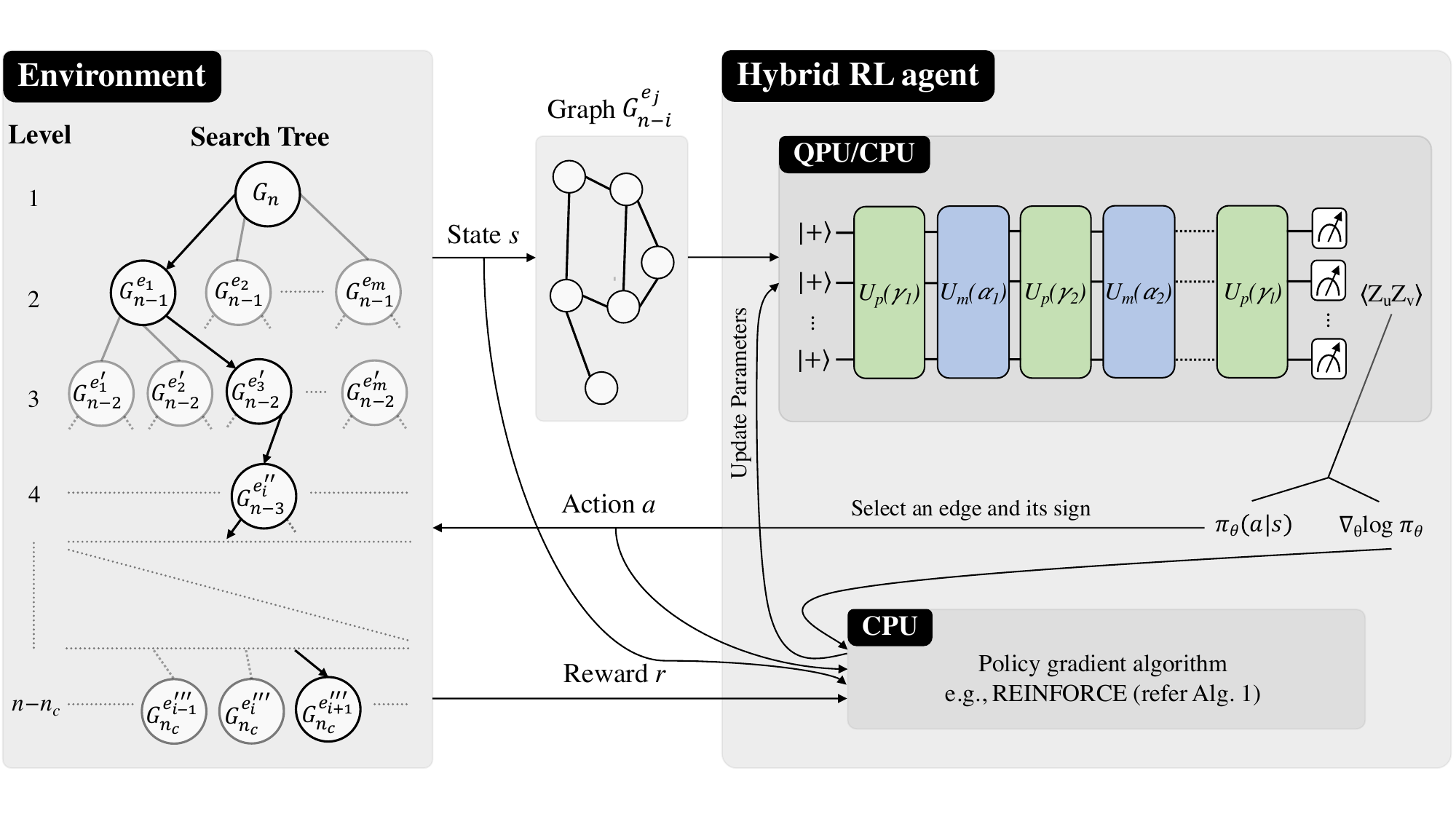}
    \caption{\textbf{Training QAOA-based policies for reinforcement learning.} We consider an RL-enhanced recursive QAOA (RL-RQAOA)  scenario where a hybrid quantum-classical agent learns by interacting with an environment which we represent as a \emph{search tree} induced by the recursive framework of RQAOA. 
    The agent samples the next action \(a\) (corresponding to selecting an edge and its sign) from its policy \(\pi_{\theta}(a|s)\) and receives feedback in the form of a reward \(r\), where each state corresponds to a graph (the state space is characterized by a search tree of weighted graphs, where each node of the tree corresponds to a graph). 
    The nodes at each level of the search tree correspond to the candidate states for an agent to perceive by taking action. 
    For our hybrid agents, the policy \(\pi_{\theta}\) of RL-RQAOA (see Def.~\protect{\ref{def:softmax_qaoa}}) along with the gradient estimate \(\nabla_{{\theta}} \log \pi_{{\theta}}\) is evaluated on a CPU as we are in the regime where depth \(l=1\). 
    However, the policy can also be evaluated on a quantum processing unit (QPU) for higher depths, when classical simulations can only be performed efficiently for graphs of small size. The training of the policy is performed by a classical algorithm such as \(\mathsf{REINFORCE}\) (see Alg.~\protect{\ref{alg:reinforce}}), which uses sample interactions and policy gradients to update parameters.}
    \label{fig:big_picture}
\end{figure*}

\subsection{Classical Simulatability of QAOA for the Ising problem}\label{subsec:classical_sim}

Next, we review the classical simulatability of a paradigmatic case of QAOA for the Ising problem. 
This is a core building block for simulating both (depth-1) RQAOA and RL-RQAOA. It enables their efficient classical simulation at depth-1 for arbitrary graphs.  
Given a graph \(G_n = (V,E)\) with \(n\) vertices \(V = [n]\) (where \([n] = \{1, 2,\ldots, n\}\)) and edges \(E \subset V \times V\), as well as an external field \(h_u \in \mathbb{R}\) and a coupling coefficient (edge weight) \(J_{uv} \in \mathbb{R}\) associated with each vertex and edge respectively, then the Ising problem aims to find a spin configuration \(s \in\{-1, +1\}^n\) maximizing the cost Hamiltonian\footnote{The textbook Ising problem definition has a negative sign \((-)\), and the goal is to minimize the Hamiltonian.},
\begin{eqnarray}
    H_n = \sum_{u \in V} h_u Z_u + \sum_{(u,v) \in E} J_{uv} Z_u Z_v.
    \label{eq:ising_func}
\end{eqnarray}
The Ising problem without any external field is equivalent to the weighted Max-Cut problem, where the goal is to find a bi-partition of vertices such that the total weight of the edges between them is maximized.
The expected value of each Pauli operator \(Z_u\) and \(Z_uZ_v\) on depth-1 QAOA can be computed classically in \(O(n)\) time using analytical results stated in Theorem~\ref{thm:analytical_form} in Appendix~\ref{app:classical_sim_qaoa}. 
Since the cost function has \(O(n^2)\) many terms in the worst case, computing the final expected value of (\ref{eq:ising_func}) hence takes a total time in \(O(n^3)\) given the variational parameters.

\subsection{Recursive QAOA}\label{subsec:rqaoa}
In this subsection, we outline the RQAOA algorithm of Bravyi et al.~\cite{bravyi2020obstacles} for the Ising problem as defined in (\ref{eq:ising_func}) with no external fields \((h_u = 0, \forall u \in V)\). 
This will serve as a base for our proposal of RL-enhanced RQAOA.
The RQAOA algorithm aims to approximate the maximum expected value\footnote{The bitstring \(\{0,1\}^n\) is analogous to the spin configuration \(\{-1,+1\}^n\) where \(0\) corresponds to \(-1\) and +1 to \(1\). Hereafter, we will use both of them interchangeably.}  \(\max_x \bra{x} H_n \ket{x}\), where \(x \in \{0, 1\}^n\).
It consists of the following steps. 
First, a standard depth-\(l\) QAOA is executed to find the quantum state \(\ket{\Psi^*_l(\vec{\alpha}, \vec{\gamma})}\) (with optimal variational parameters) as in (\ref{eq:variationalstate}) that maximizes the expectation value of \(H_n\). 
 For each edge \((u, v) \in E\), the two-correlation \(M_{u, v} = \bra{\Psi^*_l(\vec{\alpha}, \vec{\gamma})} Z_uZ_v \ket{\Psi^*_l(\vec{\alpha}, \vec{\gamma})}\) is computed. 
A variable \(Z_u\) with largest \(|M_{u, v}|\) is then eliminated (breaking ties arbitrarily) by imposing the constraint 
\begin{equation}
    Z_u = \mathrm{sign}(M_{u, v})Z_v
    \label{eq:constraint_rqaoa}
    \end{equation}
which yields a new Ising Hamiltonian \(H_{n-1}\) with at most \(n-1\) variables. 
The resulting Hamiltonian is processed iteratively, following the same steps. Finally, this iterative process stops once the number of variables is below a predefined threshold \(n_c\). 
The remaining Hamiltonian with \(n_c\) variables can then be solved using a classical algorithm (e.g., brute force method). 
The final solution can then be obtained iteratively by reconstructing eliminated variables using (\ref{eq:constraint_rqaoa}). 

 We note that the variable elimination scheme in RQAOA is analogous to rounding solutions obtained by solving continuous relaxations of combinatorial optimization problems. 
 We refer the interested reader to~\cite[Sec. \RomanNumeralCaps{5}.A.]{mcclean2021low} for a detailed discussion on the connection between quantum optimization algorithms and classical approximation algorithms. 
 Recall that the final expected value of \(H_n\) as in (\ref{eq:ising_func}) can be computed in \(O(n^3)\) time. Since we can choose \(n_c\) such that \(n_c \approx O(1)\), RQAOA runs for approximately \(n\) iterations, so that the total running time is \(O(n^4)\) (neglecting the running time needed to select the variational parameters).

\subsection{Reinforcement Learning Primer}\label{subsec:rl_primer}

As our proposal to improve upon RQAOA is based on reinforcement learning, we introduce basic concepts behind RL and the policy gradient method in this subsection. 

In RL, the agent learns an optimal policy by interacting with its environment using a trial-and-error approach~\cite{sutton2018reinforcement}. 
Formally, RL can be modeled as a Markov Decision Process (MDP) defined by the tuple \((\mathcal{S}, \mathcal{A}, p, R)\), where \(\mathcal{S}\) and \(\mathcal{A}\) represent the state and action spaces (both can be continuous and discrete), the function \(p: \mathcal{S} \times \mathcal{S} \times \mathcal{A} \rightarrow [0,1]\) defines the transition dynamics, and \(R: \mathcal{S} \times \mathcal{A} \rightarrow \mathbb{R}\) describes the reward function of the environment.
An agent's behaviour is governed by a stochastic policy \(\pi_{\theta}(a|s) : \mathcal{S} \times \mathcal{A} \rightarrow [0, 1]\), for \(a \in \mathcal{A}\) and \(s \in \mathcal{S}\).
Highly expressive function approximators, such as deep neural networks (DNN), can be used to parametrize a policy \(\pi_{\theta}\) using tunable parameters \(\theta \in \mathbb{R}^d\).
An agent's interaction governed by a policy \(\pi_{\theta}(a|s)\) in the environment can be viewed as sampling a trajectory \(\tau \sim \mathbb{P}_E(\cdot)\) from the MDP, where \(\mathbb{P}_E(\tau) = p_0(s) \pi_{\theta}(a|s) p(s_1|s,a) \cdots \pi_{\theta}(a_{H-1}|s_{H-1}) p(s_H|s_{H-1},a_{H-1})\) is the probability of the trajectory \(\tau\) of length \(H\) to occur, where \(p_0\) is a distribution of initial state \(s\).
An example of a trajectory is \(\tau = (s, a, s_1, a_1, \ldots, s_{H-1}, a_{H-1}, s_{H})\). An agent collects a sequence of rewards based on its interactions with the environment. 
The metric that assesses an agent's performance is called the value function \(V_{\pi_{\theta}}\) and takes the form of a discounted sum as follows,
\begin{eqnarray}\label{eqn:value_func}
V_{\pi_{\theta}}(s) = \mathbb{E}_{\pi_{\theta}, \mathbb{P}_E} \left(\sum_{t=0}^{H-1} \gamma^t r_t \right)  = \mathbb{E}_{\pi_{\theta}, \mathbb{P}_E} \left(R(\tau)\right)
\end{eqnarray}
where \(s\) is an initial state of an agent's trajectory \(\tau\) within an environment, \(\mathbb{P}_E\) describes the environment dynamics (i.e., in the form of an MDP)), and \(r_t\) is the reward at time step \(t\) during the interaction. 
Every trajectory has a horizon (length) \(H \in \mathbb{N} \cup \{\infty\}\) and the expected return involves a discounting factor \(\gamma \in [0,1]\). Most often one chooses \(\gamma < 1\) to avoid unwanted diverging value functions for a horizon \(H = \infty\). 
Finally, the goal of an RL algorithm is to learn an \emph{optimal policy} \(\pi^*_{\theta}\) such that the value function is maximized for each state. 
One way of finding a good policy is through the policy gradient method, i.e., finding an optimal set of parameters \(\theta\) which maximize the value function of the policy (by evaluating its gradient).
For the sake of brevity, we defer the explanation of the policy gradient method to Appendix~\ref{app:pgm}.

\section{Related Work}\label{sec:related_work}

In the context of RL, two works~\cite{yao2020policy,sung2020using} developed optimizers based on policy gradient methods for VQA optimization, highlighting the robustness of RL-based techniques against off-the-shelf optimizers in the presence of noise.
As opposed to our work, both these works use an external RL policy to choose the angles of QAOA in a one-step Markov Decision Process (MDP) environment, and otherwise rely on the basic QAOA algorithm.
A series of works~\cite{yao2021reinforcement, yao2021rl} have also used RL-based optimization to generalize the approach of QAOA for preparing the ground state of quantum many-body problems.
In~\cite{yao2021reinforcement}, an agent uses an auto-regression mechanism to sample the gate unitaries in a one-step MDP and employs an off-the-shelf optimizer to optimize angles to prepare a generalized QAOA ansatz. 
The same set of authors then unify their previous works~\cite{yao2020policy, yao2021reinforcement} with both the use of a generalized autoregressive architecture that incorporates the parameters of the continuous policy and an extended variant of Proximal Policy Optimization (PPO) applicable to hybrid continuous-discrete policies~\cite{yao2020noise}. 
We note that for all the works~\cite{yao2020policy, yao2020noise, yao2021reinforcement, yao2021rl}, the quantum circuit (QAOA-type ansatz) is a part of an environment.
In our case, we focus on employing reinforcement learning to enhance the performance of the RQAOA, inspired by a recent work~\cite{jerbi2021parametrized} on using quantum circuits to design RL policies. 
In contrast to the approaches discussed above, we design an RL policy based on QAOA ansatz in a multi-step MDP environment where the quantum circuit (QAOA ansatz) is \emph{not} a part of the environment. 
Other works have used Q-learning to formulate QAOA into an RL framework to solve difficult combinatorial problems~\cite{wauters2020reinforcement} and in the context of digital quantum simulation~\cite{khairy2020learning}. 

In the context of employing non-local post-processing methods in quantum optimization algorithms akin to classical iterated rounding, there have been a few proposals to modify RQAOA. 
The main idea behind RQAOA is to use QAOA iteratively to compute correlations and then, at every iteration, employ a rounding (variable elimination) procedure to reduce the size of the problem by one. 
The variants of RQAOA proposed in the literature primarily differ in how the correlations are computed and how the variables are eliminated. 
For instance, in~\cite{bravyi2020obstacles, bravyi2020hybrid}, variable elimination scheme of RQAOA is deterministic and relies on correlations between qubits (qudits). On the other hand, the authors in~\cite[Sec. \RomanNumeralCaps{5}.A.]{mcclean2021low} propose a modified RQAOA where the rounding procedure is stochastic (controlled by a fixed hyper-parameter \(\beta\)), and a variable is eliminated based on individual spin polarizations. 
In contrast, our proposal of RL-RQAOA trains analogous parameter(s) \(\vec{\beta}\) via RL (See Appendix~\ref{app:implementation}) and uses correlations between qubits to perform variable elimination.

\noindent \emph{Note Added:} Several pre-prints on iterative/recursive quantum optimization algorithms generalizing RQAOA have appeared since the submission of this work on arXiv. 
Parallel works such as~\cite{brady2023iterative, finvzgar2023quantum, dupont2023quantum} widen the selection and variable elimination schemes within the framework of recursive quantum optimization in application to constrained problems such as Maximum Independent Set (MIS) and Max-$2$-SAT. Moreover,~\cite{brady2023iterative} show theoretical justifications of why depth-1 QAOA might not be a suitable candidate for quantum advantage and consequently urge the community to explore higher depth alternatives.

\section{Limitations of RQAOA}\label{sec:rqaoa_lim}

This section highlights some algorithmic limitations of RQAOA by introducing an alternative perspective on it. Then, based on this perspective, we provide insights into when RQAOA might fail and why.
It is obvious that (depth-1) RQAOA must fail on some instances, since we assume \(\mathsf{BPP} \subsetneq \mathsf{NP}\)\footnote{Here, we use the complexity-theoretic assumption of \(\mathsf{BPP} \subsetneq \mathsf{NP}\) because depth-1 RQAOA can be simulated classically.}, but these instances may be quite big a priori.
By ``failure'', we mean that RQAOA can not find an optimal (exact) solution. 
Notably, even if depth-$l$ RQAOA fails to find exact solutions, it could still achieve an approximation ratio better than the bound known from inapproximability theory. 
In this case, $NP \subseteq BQP$ still holds.
For instance, if RQAOA fails to find an exact solution and still achieves an approximation ratio of $16/17 + \epsilon$ or $0.8785 + \epsilon$, then $NP \subseteq BQP$ follows from~\cite{haastad2001some, khot2007optimal} under different complexity-theoretic assumptions, thus demonstrating quantum advantage.
We primarily focus on finding small-size instances since we need a data set of small instances to be able to computationally efficiently compare the performance of (depth-1) RQAOA and RL-RQAOA.

\begin{figure*}
    \centering
    \includegraphics[width=2\columnwidth]{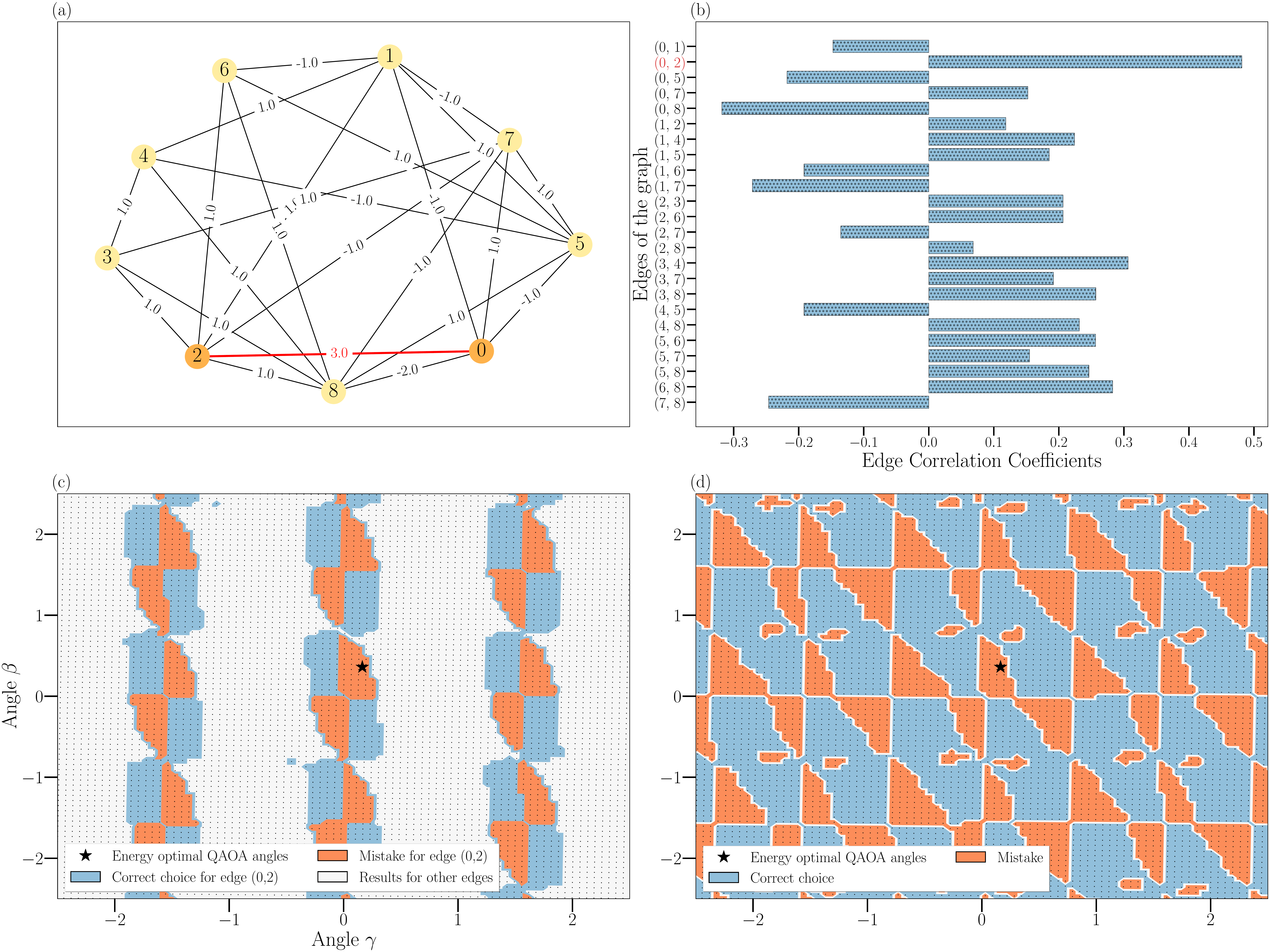}
    \caption{\textbf{Illustration of a counterexample where the heuristic of using the energy-optimal QAOA angles in RQAOA fails.} Here, we show that for the weighted graph (9 vertices and 24 edges) depicted in (a), RQAOA makes a mistake even in its strongest regime, so at the very first iteration (i.e., \(n_c = 8\)). The two-correlation coefficients for each edge (at energy-optimal angles) are shown in the form of a horizontal bar plot in (b), where the edge \((0,2)\) has the maximal correlation coefficient. For the graph in (a), RQAOA with energy-optimal angles assigns a wrong edge-correlation (sign) to this edge which is precisely highlighted by a bold star in (c) and (d). Both (c) and (d) characterize the sets of good and bad QAOA angles where RQAOA makes a correct and a wrong choice, respectively.
    This example is counter-intuitive: as the edge \((0,2)\) has the highest weight in the graph, intuitively, the variables should be correlated (same sign) as to maximize the energy. However, this leads to a sub-optimal solution which RQAOA achieves with energy-optimal angles. Yet, for different settings of QAOA angles which do not maximize the overall energy, this edge will still have the largest magnitude of correlation, but in this case, anti-correlation, which leads to the true optimum (see sub-figure (c)). 
    }
    \label{fig:couterexample}
\end{figure*}

First, let us motivate the use of QAOA as a subroutine in RQAOA. In other words, why would one optimize depth-\(l\) QAOA (i.e., find energy-optimal parameters for a Hamiltonian) and then use it in a completely different way (i.e., perform variable elimination by computing two-correlation coefficients \(M_{u,v}\)). 
Intuitively, using QAOA in such a fashion makes sense because as depth \(l \rightarrow \infty\), the output of QAOA converges to the quantum state which is the uniform superposition over all optimal solutions, and hence, for each pair \((u,v) \in E\), computing the coefficient \(M_{u,v}\) \emph{exactly} predicts if the edge is correlated (vertices with the same sign; i.e. lie in the same partition) or anti-correlated (vertices with the different sign; i.e. lie in the different partition) in an optimal cut. 
The next piece of intuition, which is not any kind of a formal argument, is that low-depth QAOA prepares a superposition state where low-energy states are more likely to have high probability amplitudes.
Then the RQAOA selects the edge which is most correlated or anti-correlated in these low-energy states. 
Furthermore, assuming that an ensemble of reasonable solutions often agree on which edges to keep and which ones to cut, RQAOA will select good edges to cut or keep (from the Max-Cut perspective).
However, we also expect RQAOA to fail sometimes, for instance, when the intuition mentioned above is wrong, or it assigns a wrong edge-correlation sign to an edge for other reasons. 
Hence, as RQAOA fails, this raises the question of whether there are better angles to select an edge and its correct edge-correlation sign at every iteration than those which coincide with energy-optimal angles (see Fig.~\ref{fig:couterexample}).

RQAOA can alternatively be visualized as performing a tree search to find the most probable spin configuration close to the ground state of the Ising problem. 
In particular, at the \(k^{th}\) level of the tree, nodes correspond to graphs with \(n-k\) vertices, each having different edge sets. 
Suppose that a node has \(n-k\) vertices with \(e\) edges, then it will have \(e\) many children where each child corresponds to a graph with \(n-k-1\) vertices having different edge sets following the edge contraction rules by imposing (\ref{eq:constraint_rqaoa}). 
The original RQAOA proposal~\cite{bravyi2020obstacles} is a randomized algorithm (in the sense that ties between maximal two-correlation coefficients are broken uniformly at random) on this tree exploring only a single path during one run and terminating at the \((n-n_c)^{th}\) level. 
The decision of choosing an appropriate branch is performed based on the largest magnitude of the absolute value of two-correlation coefficients \(M_{u, v}\) computed via a depth-$l$ QAOA using \(H_n\)-energy optimal parameters. 
While exploring level-by-level, RQAOA assigns the edge correlations \((-1~\text{or}~+1)\)
where a vertex is eliminated according to the constraint (\ref{eq:constraint_rqaoa}). 
We note that in the case of ties between maximal two-correlation coefficients, independent runs of RQAOA might not necessarily induce the same search tree.

This alternative perspective described above provides some insights regarding the limitations of RQAOA: (i) when there are ties and branching occurs, it could be that only one path within a set of induced search tree leads to a good approximate solution; and (ii) it may be the case that even when there are no ties (i.e., one path and no branching), selecting edges to contract according to the maximal correlation coefficient stemming from energy-optimal parameters of QAOA is an incorrect choice to attain a good solution. 
A priori, it is not obvious if any of the above mentioned two possibilities can occur under the choice of energy-optimal angles. 
However, note that one of (i), (ii), or a combination of both must happen; otherwise, RQAOA is an efficient polynomial-time algorithm for the Ising problem. 
Hence, in the case that RQAOA makes an incorrect choice, RQAOA lacks the ability to explore the search tree to find better approximate solutions. 
Keeping these considerations in mind, we will show later that both phenomena (i) and (ii) occur by performing an empirical analysis of RQAOA.
We now describe both the limitations mentioned above in detail below.

\begin{figure}[htb]
    \centering
    \includegraphics[width=\columnwidth]{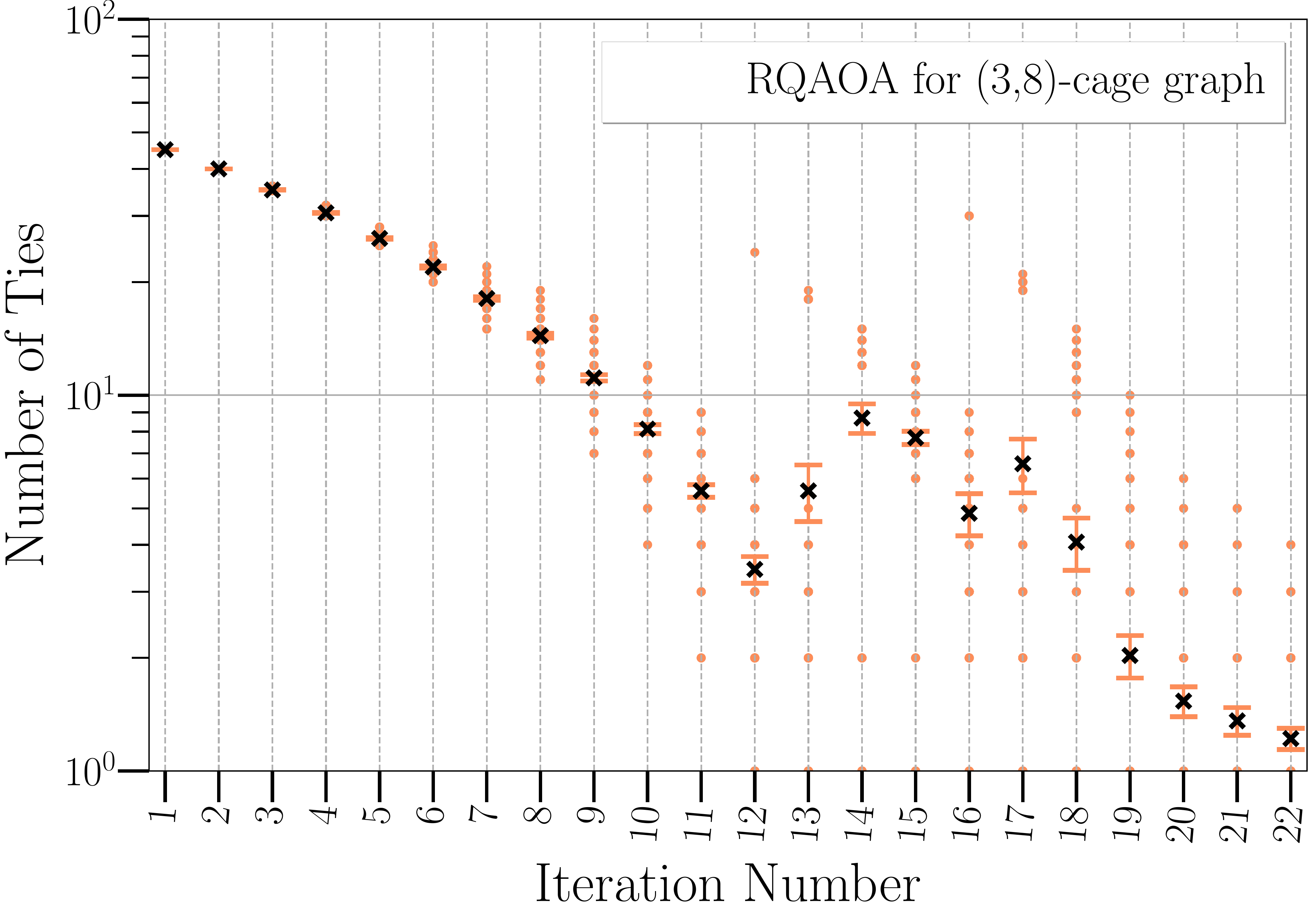}
    \caption{Number of ties per iteration of RQAOA (average over 200 runs) for \((3,8)\)-cage graph (\(30\) vertices, \(45\) edges, edge weights \((\{-1, +1\})\). We chose \(n_c=8\) in our simulations where RQAOA achieved a mean approximation ratio of \(0.955 \pm 0.036\) and the probability to reach the ground state was \(33.5\%\). The y-axis (Number of Ties) is log-scaled.
    The black crosses depict the mean values, with the error bar showing the 95\% confidence interval of 200 independent runs.
    The figure illustrates that one would invariably encounter a constant fraction of ties between maximal two-correlations no matter whatever path is chosen in the search tree, implying an exponential blow-up in the size of the search tree to be explored by RQAOA.}
    \label{fig:cage_ties}
\end{figure}

\begin{itemize}
    \item[(i)]\label{item:ties} It may be the case that eliminating a variable by taking the 
    \emph{argmax} of the absolute value of two-correlation coefficients is always a correct choice, but there can be more than one choice at every iteration.
    Moreover, it is possible to construct instances with a small number of optimal solutions, where for the majority of \(n-n_c\) iterations (corresponding to the level of the tree) there is at least one tie (here, \(m\) ties corresponds to \(m+1\) pairs \((u_1, v_1), \ldots, (u_{m+1}, v_{m+1})\) with the same two-correlation coefficient).
    In other words, the number of times RQAOA needs to traverse the search tree in the worse case to reach the ground state (optimum) may be exponentially large; i.e., every argmax tie break leads to a new branching of the potential choices of RQAOA, and this happens at each level of the tree.
    We showcase this phenomenon in our empirical analysis for one such family of instances (see Fig.~\ref{fig:cage_ties}).
    One may imagine perturbing the edge weights to avoid ties while preserving the ground states of the Hamiltonian, but no such perturbation is generally known. 
    
    \item[(ii)]\label{item:rqaoa_mistake} It may be the case that the path to reach the ground state requires the selection of a pair \((u,v)\) (and its correlation sign) for which the two-correlation coefficient is \emph{not} maximal according to QAOA at energy-optimal parameters (see Fig.~\ref{fig:couterexample}).
    This implies that RQAOA might be prematurely locking out on optimal solutions. 
    
\end{itemize}

We provide examples of graphs to prove the validity of the observations above.
In the regime where there are ties between maximal correlation coefficients [(i)], we performed 200 independent RQAOA runs for the family of weighted \((d,g)\)-cage graphs\footnote{A \((d,g)\)-cage graph \((d \geq 3, g \geq 5)\) is a \(d\)-regular graph of girth \(g\) (length of a shortest cycle contained in the graph)  consisting of the smallest number of vertices possible.}
\((3 \leq d \leq 7; 5 \leq g \leq 12;~\text{edge weights}~\{-1, +1\})\) where ties are broken uniformly at random for the \(n-n_c\) iterations (levels of the tree). 
We work with these graphs because the subgraphs that (depth-1) QAOA sees are regular trees (for most edges at every iteration of RQAOA, QAOA will see a \((d-1)\)-ary tree, as cage graphs are \(d\)-regular graphs, which creates the situation of ties between correlation coefficients).
Here, by \emph{seeing} we refer to the fact that the output of depth-\(l\) QAOA for a qubit (vertex) only depends on the neighbourhood of qubits that are within \(l\) distance to the given qubit~\cite{farhi2020mc}.
For these graphs, we found that in \(86.4 \pm 9.63\%\) of the \(n-n_c\) iterations, the variable to eliminate was chosen from the ties between maximal correlation coefficients (see Fig.~\ref{fig:cage_ties}).

To investigate the scenario of [(ii)], we focus on a particular case where there are no ties (or comparatively less ties) and find instances such that taking the maximal two-correlation coefficient does not reach the optimum solution in the tree. For this, we performed a random search over an ensemble of \(10600\) weighted random \(d\)-regular graphs and found several small-size instances (\#nodes \(\leq 30\)) for which RQAOA did not attain the optimum.

Using both the theoretical and numerical observations discussed above, we create a dataset of graphs (containing both hard and random instances for RQAOA) for our later analysis. 
In the next section, we develop our new algorithm (RL-RQAOA) and compare its performance to RQAOA to assess the benefit of employing reinforcement learning in the context of recursive quantum optimization specifically for hard instances. 
Finally, we give the relevant details about the data set of the graph ensemble considered in Sec.~\ref{subsec:hard_inst}.

\section{Reinforcement Learning Enhanced RQAOA \& Classical Brute Force Policy}\label{sec:rl_rqaoa}

Having introduced the background of policy gradient methods and the limitations of RQAOA, we develop a QAOA-inspired policy which selects a branch in the search tree (eliminate a variable) at every iteration of RL-RQAOA. 
Recall that, even though selecting an edge to contract according to the maximal two-correlation coefficient is often a good choice, it is not always an optimal one, and also often, there is no single best option, but more (for instance, see Fig.~\ref{fig:couterexample}).
Our basic idea is to train an RL method to learn how to select the edges to contract (along with its edge-correlation sign) correctly while using the information generated by QAOA.
Additionally, to investigate the power of the quantum circuit within the quantum-classical arrangement of RL-RQAOA, we design a classical analogue of RL-RQAOA called reinforcement-learning recursive ONE (RL-RONE) and compare it with RL-RQAOA. 

To overcome the limitations of RQAOA, one needs to carefully tweak (a) RQAOA's variable elimination subroutine and (b) the use of QAOA as a subroutine; i.e., instead of finding energy-optimal parameters, we learn the parameters of QAOA. For (a), we apply the non-linear activation function \(\mathsf{softmax}_{\vec{\beta}}\) (see Def.~\ref{def:softmax_qaoa}) on the absolute value of two-correlation coefficients \(|M_{u,v}|\) measured on \(\ket{\Psi_l(\vec{\alpha}, \vec{\gamma})}\). 
By doing this, the process of selecting a variable to eliminate (and its sign) is represented by a smooth approximation of \emph{argmax} that is controlled by a vector of \emph{trainable} inverse temperature parameters \(\vec{\beta}\) (one \(\beta\) per edge).
The parameters \(\vec{\beta}\) (initialized at low values) are then trained such that the probability of selecting an edge (or a branch at every iteration) with the highest expected reward tends to 1.
In the case of (b), we train the variational angles of QAOA in the course of learning rather than using the ones that give optimal energy. 
We do this because of the following two reasons: (i) to avoid costly optimization\footnote{We train the QAOA angles at depth-1 even though we can optimize them efficiently (see Appendix~\ref{app:var_opt}). However, the optimization becomes non-trivial with an increase in depth.}; (ii) different angle choices can help the algorithm sometimes to choose optimal paths in the search tree that are not possible otherwise (see Fig.~\ref{fig:couterexample}).
We note that the entire learning happens on one instance of the Ising problem.
Even though it is conceivable to train the algorithm over an ensemble of instances by introducing suitable generalization mechanisms such that \(\vec{\beta}\) are dependent on instances, we solely focus on learning parameters of the policy of RL-RQAOA on one instance so that it eventually performs better than RQAOA. \hfill \break

\begin{algorithm*}[H]\label{alg:reinforce}
\DontPrintSemicolon

  \SetAlFnt{\small\sffamily}

  \SetNlSkip{1em}
  
  \KwInput{The policy of RL-RQAOA (Def.~\ref{def:softmax_qaoa}) or RL-RONE (Def.~\ref{def:softmax_one})}
    \BlankLine
Initialize the policy parameters \(\theta\).\;
\While{True}
{
     Generate \(N\) episodes \(\left\{(s_0, a_0, r_1, \ldots, s_{H-1}, a_{H-1}, r_H)\right\}_i\) following \(\pi_{\theta}\).\;
     
     \For{episode i in batch}{
        Compute the returns \(G_{i,t} \leftarrow \sum_{t'=1}^{H-t}\gamma^{t'} r_{t+t'}^{(i)}\).\;
        
        Compute the gradients \(\nabla_{{\theta}} \log \pi_{{\theta}}(a_{t}^{(i)}| s_{t}^{(i)})\).\;
     }
    
    Compute \(
\Delta{\theta}=\frac{1}{N} \sum_{i=1}^{N} \sum_{t=0}^{H-1} \nabla_{{\theta}} \log \pi_{{\theta}}(a_{t}^{(i)}| s_{t}^{(i)}) \ G_{i, t}\).   \;   

    Update \(\theta \leftarrow \theta + \delta \Delta\theta\).

}
       
\caption{\(\mathsf{REINFORCE}\) algorithm for the policies of RL-RQAOA and RL-RONE}
\end{algorithm*} \hfill \break

To provide further details on the effective Markov Decision Process (MDP) that the above described policy will be exploring, note that the RQAOA method can be interpreted as a multi-step (also called a \(n\)-step) MDP environment (with a delayed reward and a non-trainable policy), where at every iteration, a variable is eliminated based on the information generated by QAOA. Let us now cast the learning problem of variable elimination in the RL framework, inspired by recent work~\cite{jerbi2021parametrized} on using quantum circuits to design RL policies.
For every step of the episode\footnote{Here, one episode corresponds to one complete run of (RL)-RQAOA.}, our RL agent is required to choose one action out of the discrete space equivalent to an edge set of the underlying graph; i.e., in the worse case, selects one edge from \(n \choose 2\), on which it imposes a constraint of the form (\ref{eq:constraint_rqaoa}).
Hence, the state space \(\mathcal{S}\) consists of weighted graphs (which we could encounter during an RQAOA run) and the action space \(\mathcal{A}\) consists of edges (and $\pm 1$ edge-correlations to impose on them).
The actions are selected using a parameterized policy \(\pi_{\theta}(a|s)\) which is based on the QAOA ansatz.
Since, we use the expectation value of the Hamiltonian \(H_n\) of the Ising problem as an objective function, the reward space is \(\mathcal{R} = [0, \max_{x \in \{0,1\}^n} \bra{x}H_n\ket{x}]\). 
 
Next, we formally define the policy of RL-RQAOA and its learning algorithm, which is a crucial part of RL-RQAOA.  

\begin{defn}[Policy of RL-RQAOA]\label{def:softmax_qaoa}
Given a depth-l QAOA ansatz acting on \(n\) qubits, defined by a Hamiltonian \(H_n\) (with an underlying graph \(G_n = (V,E)\)) and variational parameters \(\{{\vec{\alpha}}, \vec{{\gamma}}\} \in [0, 2\pi]^{2l}\), let \(M_{u, v} = \bra{\Psi_l(\vec{{\alpha}}, \vec{{\gamma}})} Z_u Z_v \ket{\Psi_l(\vec{{\alpha}}, \vec{{\gamma}})}\) be the two-correlations that it generates. We define the policy of RL-RQAOA as
\begin{equation}\label{eqn:softmax_policy}
\pi_{\theta}(a = (u, v)|s = G_{n})=\frac{\exp(\beta_{u,v} \cdot |M_{u,v}|)}{\sum\limits_{(u,v) \in E} \exp(\beta_{u,v} \cdot |M_{u,v}|)}
\end{equation}
where actions \(a\) correspond to edges \((u,v) \in E(G_n)\), states \(s\) to graphs \(G_n\) and \(\beta_{u,v} \in \mathbb{R}\) (exists for every possible edge) is an inverse temperature parameter. Here, \(\theta = (\vec{\alpha}, \vec{\gamma}, \vec{\beta})\) constitutes all trainable parameters, where \(\vec{\beta} \in \mathbb{R}^{(n^2 - n)/2}\).
\end{defn}

The reader is referred to Alg.~\ref{alg:rlrqaoa} for the pseudo-code of RL-RQAOA (for one episode), where the addition of RL components are highlighted in the shade of green. 
Furthermore, we note that RL-RQAOA is a generalized version of RQAOA because the former is exactly equivalent to the latter when the energy-optimal parameters \(\{\vec{\alpha}, \vec{\gamma}\}\) are specified by QAOA on \(H_n\), and for all \((u, v) \in E,~\beta_{u,v} \in \vec{\beta},~\text{where}~\beta_{u,v} = \infty\). 

Since the vector \(\vec{\beta}\) is edge specific and as we learn \(\beta_{u,v} \in \vec{\beta}~\text{for}~\{u,v\} \in E\) separately for every instance, we develop a fully classical RL algorithm, namely RL-RONE, to simply learn \(\beta_{u,v}\) for all edges directly in spite of where the two-correlation coefficients \(M_{u,v}\) are generated from.
It is natural to consider this because it might be the case that in the hybrid quantum-classical arrangement of RL-RQAOA, the classical part (learning of \(\beta_{u,v}~\text{for}~\{u,v\} \in E\)) is more powerful than the quantum part (computing two-correlations \(M_{u, v}~\text{for}~\{u,v\} \in E\) from the QAOA ansatz at given variational angles \(\{{\vec{\alpha}}, \vec{{\gamma}}\}\)). 
Hence, in order to assess the contribution of the quantum circuit in RL-RQAOA, we define the policy of RL-RONE
such that for each edge, we fix the two-correlation \(M_{u, v} = 1\); i.e., we do not use any output from the quantum circuit. 
Although simply using \(M_{u, v} = 1\) in Def.~\ref{def:softmax_qaoa}, the policy will select an edge and always assign it to be correlated, rendering it to be less expressive. 
A solution to this problem is to simultaneously learn the parameters \(\beta_{u,v}^{+1}~\text{(correlated edge)}~\text{and}~\beta_{u,v}^{-1}~\text{(anti-correlated edge)}\) for each edge. 
Then the resulting RL-RONE algorithm is expressive enough to reach the optimum solution. Moreover, it has trainable inverse temperature parameters \(\vec{\beta}\) where \(|\vec{\beta}| = n^2 - n\) for \(n\) the number of nodes of the graph \(G_n\).
The notion of an action slightly differs from the  RL-RQAOA policy as the action here corresponds to selecting an edge along with its sign ($+1$ and $-1$ for correlated and anti-correlated edges, respectively), while in RL-RQAOA, the two-correlation coefficient implicitly selects this sign.
We formally define the policy of RL-RONE below.
\begin{defn}[Policy of RL-RONE]\label{def:softmax_one}
Given a Hamiltonian \(H_n\) (with an underlying graph \(G_n = (V,E)\)), we define the policy of RL-RONE as 
\begin{equation}\label{eqn:softmax_one}
\pi_{\theta}(a = ((u, v), b)|s = G_{n})=\frac{\exp(\beta^b_{u,v})}{\sum\limits_{b \in \{\pm 1\}}\sum\limits_{(u,v) \in E} \exp(\beta^b_{u,v})}
\end{equation}
where actions \(a\) correspond to edges \((u,v) \in E(G_n)\) along with an edge correlation \(b \in \{\pm 1\}\), states \(s\) correspond to graphs \(G_n\) and \(\beta_{u,v}^{\pm1} \in \mathbb{R}\) (exists for every edge) are inverse temperature parameters. Here, \(\theta = (\vec{\beta}^{+1}, \vec{\beta}^{-1})\) constitutes all trainable parameters, where \(\vec{\beta}^{\pm1} \in \mathbb{R}^{(n^2 - n)/2}\). 
\end{defn}
The classical analogue RL-RONE can then be simulated by performing the following modifications to Alg. \ref{alg:rlrqaoa}: (i) modify the parameters \(\theta\) of the policy of RL-RONE by \(\theta = (\vec{\beta}^{+1}, \vec{\beta}^{-1})\), (ii) delete \(\mathsf{Lines~4}\) and \(\mathsf{5}\), (iii) update \(\mathsf{Line~6}\) by incorporating the policy of RL-RONE and the constraint (\ref{eq:constraint_rqaoa}) in \(\mathsf{Line~7}\) is imposed by feeding the correlation sign of the edge from the output (\(b \in \{\pm1\}\)) of the policy of RL-RONE. \hfill \break

We train both the policies of RL-RQAOA and RL-RONE using the Monte Carlo policy gradient algorithm \(\mathsf{REINFORCE}\), as explained in Appendix~\ref{app:pgm}. 
Also, refer to Alg.~\ref{alg:reinforce} for the pseudo-code. 
The horizon (length) of an episode is \(n-n_c\). 
The value function is defined as \(V_{\pi_{\theta}}(H_n) = \mathbb{E}_{{\pi_{\theta}}}\left[\gamma^{n-n_c} \cdot \bra{x}H_n\ket{x} \right]\), where \(\gamma \in [0,1]\), \(H_n\) is the Hamiltonian defined on \(n\) variables for a problem instance and \(x\) is a binary bitstring as defined in \(\mathsf{Line~14}\) of Alg.~\ref{alg:rlrqaoa}. 

In this work, we only focus on simulations of depth-1 RQAOA and RL-RQAOA. Indeed, the particular case of depth-1 quantum circuits and Ising Hamiltonian RQAOA can be simulated efficiently classically; see Sec.~\ref{subsec:classical_sim} and Appendix~\ref{app:classical_sim_qaoa}. 
However, classical simulatibility is not known for Ising cost functions at depth larger than 2~\cite{bravyi2021classical}, and more general cost Hamiltonians even at depth-1 (e.g., Max-\(k\)-XOR on arbitrary hypergraphs), leaving room for both quantum and RL-enhanced quantum advantage.

 \begin{algorithm*}[H]\label{alg:rlrqaoa}
\DontPrintSemicolon

  \SetAlFnt{\small\sffamily}

  \KwInput{A graph $G_n = (V, E)$ with $|V| = n$; a weighted adjacency matrix $J_n$ defined by Ising Hamiltonian $H_n$; the policy of RL-RQAOA (Def.~\ref{def:softmax_qaoa}) or RL-RONE (Def.~\ref{def:softmax_one}); $n_c$}
  \KwOutput{A binary bitstring $x \in \{\pm 1\}^n$}
    \BlankLine

{\color{black}Initialize the parameters \(\theta = (\vec{\alpha}, \vec{\gamma}, \vec{\beta})\) of the policy; \(G_{\mathrm{current}} \gets G_n\); \(H_{\mathrm{current}} \gets H_n\); \(J_{\mathrm{current}} \gets J_n\).}\;
{\color{black}Let \(\kappa^{(0)} : \{\pm 1\}^n \rightarrow \{\pm 1\}^n\) be an identity map.}\;
\For{\(i = 1\) \text{\textbf{to}} \(n-n_c\)}
{   
    
    {\color{black} Run $\mathsf{QAOA}_l(H_{\mathrm{current}})$ to find a state $\ket{\Psi_l}$ for 
     angles \(\{{\vec{\alpha}}, \vec{{\gamma}}\} \in [0, 2\pi]^{2l}\).}\; 
    
    {\color{black}Compute \(M_{u, v} = \bra{\Psi_l} Z_uZ_v \ket{\Psi_l}, \forall (u,v) \in E\).}\;

    {\color{green!55!blue} Select \((u,v) = \textsf{softmax}_{(u,v) \in E}(|M_{u,v}|, \beta_{u,v})\) as defined in (\ref{eqn:softmax_policy}).}\;
    
    {\color{black} Imposing a constraint \(Z_v = \textsf{sign}(M_{u,v})Z_u\), eliminate \(Z_v\) from \(H_{\text{current}}\) resulting into a Hamiltonian \(H_{n-i}\) with \(n-i\) variables.}\;
    
    {\color{black}Define a function \(\kappa : \{\pm1\}^{| V(G_{n-i})|}
    \rightarrow \{\pm1\}^{|V(G_{\text{current}})|}\) 
    by updating for all \(w \in V(G_{\text{current}})\)
        $ \kappa(x)_w =
    \begin{cases}
                    \textsf{sign}(M_{u,v})x_u, & \text{if } w = v.\\
                    x_w, & \text{otherwise}\\
    \end{cases}$}\;
    
    $\kappa^{(i)} \leftarrow \kappa^{(i-1)} \circ \kappa$\;
        
    {\color{black} Update \(G_{\mathrm{current}} \gets G_{n-i}\); \(H_{\mathrm{current}} \gets H_{n-i}\); and \(J_{\mathrm{current}} \gets J_{n-i}\).}\;
       }
    {\color{black} Let \(G_{n_c} = (V,E)\) be the final graph with \(|V| = n_c\) vertices.}\;
    Find \(x^* = \argmax_{x\in \{-1, 1\}^{n_c}} \bra{x}H_{n_c}\ket{x}.\)\;
    {\color{green!55!blue} Update parameters \(\theta\) using \(\mathsf{REINFORCE}\) (Alg.~\ref{alg:reinforce}).}\;
    \Return \(\kappa^{(n-n_c)}(x^*)\)\;

\caption{RL-RQAOA \((G_n= (V,E), J_n, n_c)\) for one episode}
\end{algorithm*} 

\begin{figure*}
    \centering
    \includegraphics[width=2\columnwidth]{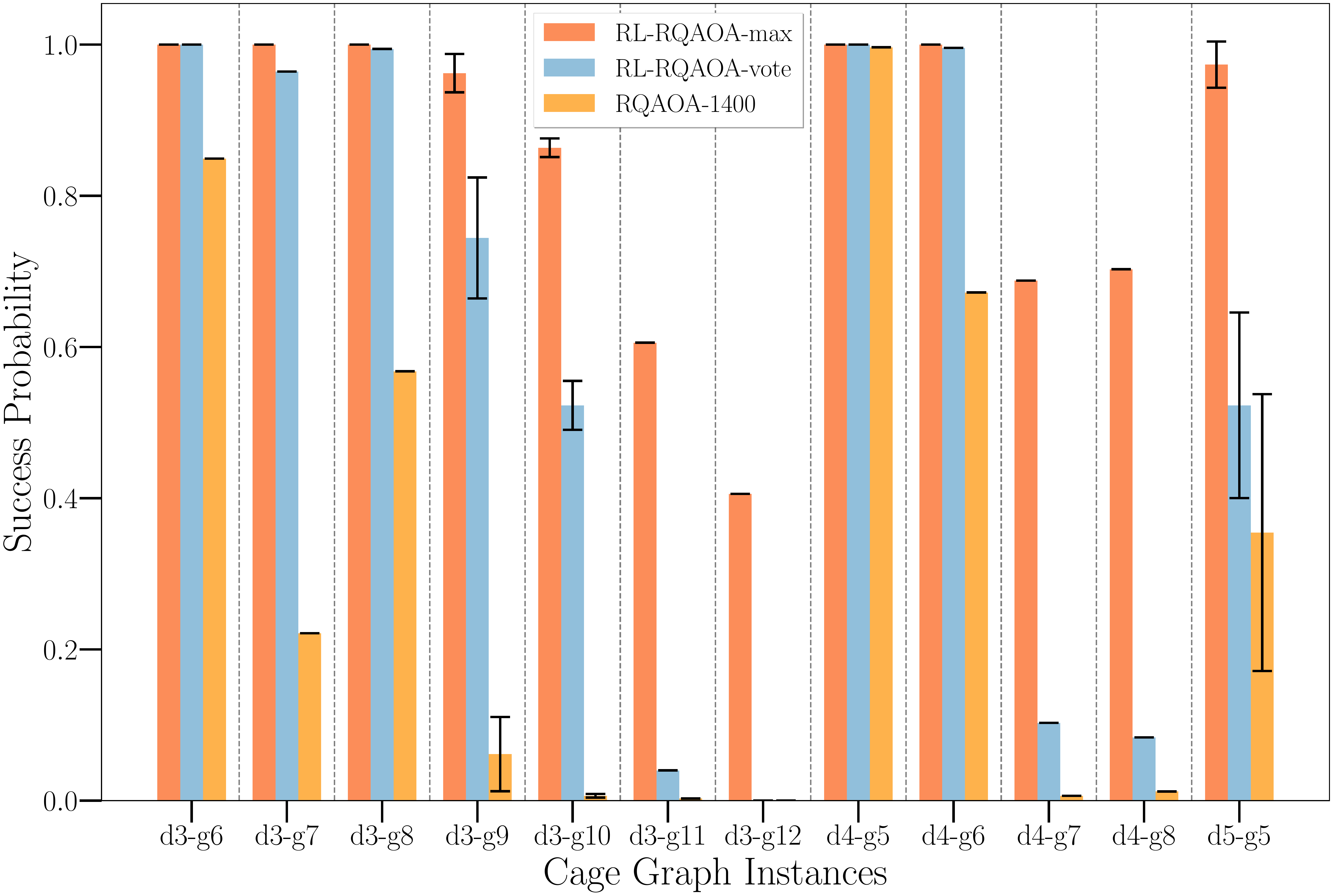}
    \caption{\textbf{Comparison of success probability in attaining ground state solutions of RL-RQAOA and RQAOA on cage graphs.} The x-axis depicts the properties of cage graph(s), for instance, d3-g6 denotes that the instance is \(3\)-regular with girth (length of the shortest cycle) being 6. 
    The error-bars appear only for few instances (specifically for d3-g9, d3-g10 and d5-g5) because of the existence of multiple graph instances with the same properties (degree and girth). The evaluation of RL-RQAOA was done by evaluating the average learning performance over 15 independent runs. While, for RQAOA, the best energy is taken when given a fixed budget of 1400 runs. The probability for RL-RQAOA-max is computed by taking the maximum energy attained by the agent over all 15 independent runs for a particular episode. One the other hand, the probability for RL-RQAOA-vote (statistically more significant) is computed by aggregating the maximum energy attained for a particular episode only if more than 50\% of the runs agree. We chose \(n_c=8\) for instances with nodes \(\leq 50\) and \(n_c=10\) otherwise. The parameters \(\theta = (\alpha, \gamma, \vec{\beta})\) of the RL-RQAOA policy were initialized by setting \(\vec{\beta} = \{25\}^{{(n^2-n)}/2}\) and the angles \(\{\alpha, \gamma\}\) (at every iteration) to energy-optimal angles (i.e., by following one run of RQAOA). All agents were trained using \(\mathsf{REINFORCE}\) (Alg. \ref{alg:reinforce}).} 
    \label{fig:rl_rqaoa_vs_rqaoa_random}
\end{figure*}

\section{Numerical Advantage of RL-RQAOA over RQAOA}\label{sec:numerics}

In the previous section, we have introduced both our quantum (inspired) policy of RL-RQAOA and an entirely classical policy of RL-RONE, and their design choices, and based on these; we propose an RL-enhanced RQAOA and its classical analogue RL-RONE.
Although we gave justifications for these choices, it is natural to evaluate their influence on the performance of RL-RQAOA and RL-RONE.
In this section, we first describe how we found hard instances for RQAOA and discuss their properties. 
We then describe the results of our numerical simulations, where we consider both hard instances and random instances to benchmark the performance of (depth-1) RQAOA, RL-RQAOA, and RL-RONE.
The reader is referred to Appendix~\ref{app:implementation} for implementation details for the above algorithms.

\subsection{Hard Instances for RQAOA}\label{subsec:hard_inst}

Here, our focus is on finding \emph{small-size hard instances} (with approximation ratio as a metric) for the Ising problem where RQAOA fails. 
Note that, we assume it must fail to solve exactly as if it does not, then \(\mathsf{NP} \subseteq \mathsf{BQP}\) as the Ising problem is \(\mathsf{NP}\)-hard in general.
As we lack techniques to analyze the performance guarantees of RQAOA at arbitrary depth \(l\) apart from special cases like ``ring of disagrees" at depth-1~\cite{bravyi2020obstacles}, it is a non-trivial task to find hard instances for RQAOA. 
In this spirit, we generate an ensemble \(\mathcal{G}[n, d, w]\) of \emph{weighted random d-regular instances} with \(n\) vertices and edge weight distribution \(w: E \rightarrow \mathbb{R}\).
We then perform a random search over \(\mathcal{G}[n, d, w]\) to find hard instances.
Concretely, we construct a graph ensemble \(\mathcal{G}[n,d,w]\) as follows: for each tuple of parameters \((n, d, w) \in \{14, 15, \ldots, 30\} \times \{3, 4, \ldots, 29\} \times \{\mathrm{Gaussian}, \mathrm{bimodal}\}\), we generate 25 graphs whenever possible\footnote{For generating \emph{d}-regular graphs with \(n\) vertices, \(1 \leq d \leq n-1\) and further if \(d\) is odd, \(n\) must be even.} 
yielding \(10600\) graphs in total, where Gaussian \((\mathcal{N}(0,1))\) and bimodal \((\{\pm1\})\) are edge weight distributions.
Intuitively, the instances with bimodal edge weights would have a huge level of degeneracy within the ground states, which is confirmed by our simulations. Moreover, for the instances with bimodal edge weights, where ties between two-correlation coefficients were encountered, the final approximation ratio was computed based on the best energy attained by running RQAOA for a maximum of 1400 independent runs.
On the other hand, for the instances with Gaussian edge weights \(\mathcal{N}(0,1)\), we found that all instances had unique ground states. Hence, we ran RQAOA only once to get the best approximation ratio for instances with Gaussian edge weights.

We filter out \(1027\) (857 with bimodal weights and 170 with Gaussian weights) instances for which RQAOA's approximation ratio is less than \(0.95\). 
Note that RQAOA can only be closer to optimal the larger \(n_c\) is. 
In other words, it monotonically improves the quality of the solution with an increase in \(n_c\).
Since we want to improve upon RQAOA in its strongest regime, we choose \(n_c = 8\) (unless specified otherwise) for our numerical simulations.
However, interestingly for the \(1027\) hard instances found above, even with \(n_c = 4\), we only found 26 instances (5 with bimodal weights and 21 with Gaussian weights) for which the approximation ratio decreased (for the rest, the approximation ratio remained the same). We chose \(n_c=4\) for the previously mentioned experiment because, for some instances, the edge weights cancelled out after an edge contraction subroutine, and as a consequence, the intermediate graph ended up being an empty graph (a graph with zero edge weights) for \(1 \leq n_c < 4\).

 \begin{figure}
    \centering
    \includegraphics[width=\columnwidth]{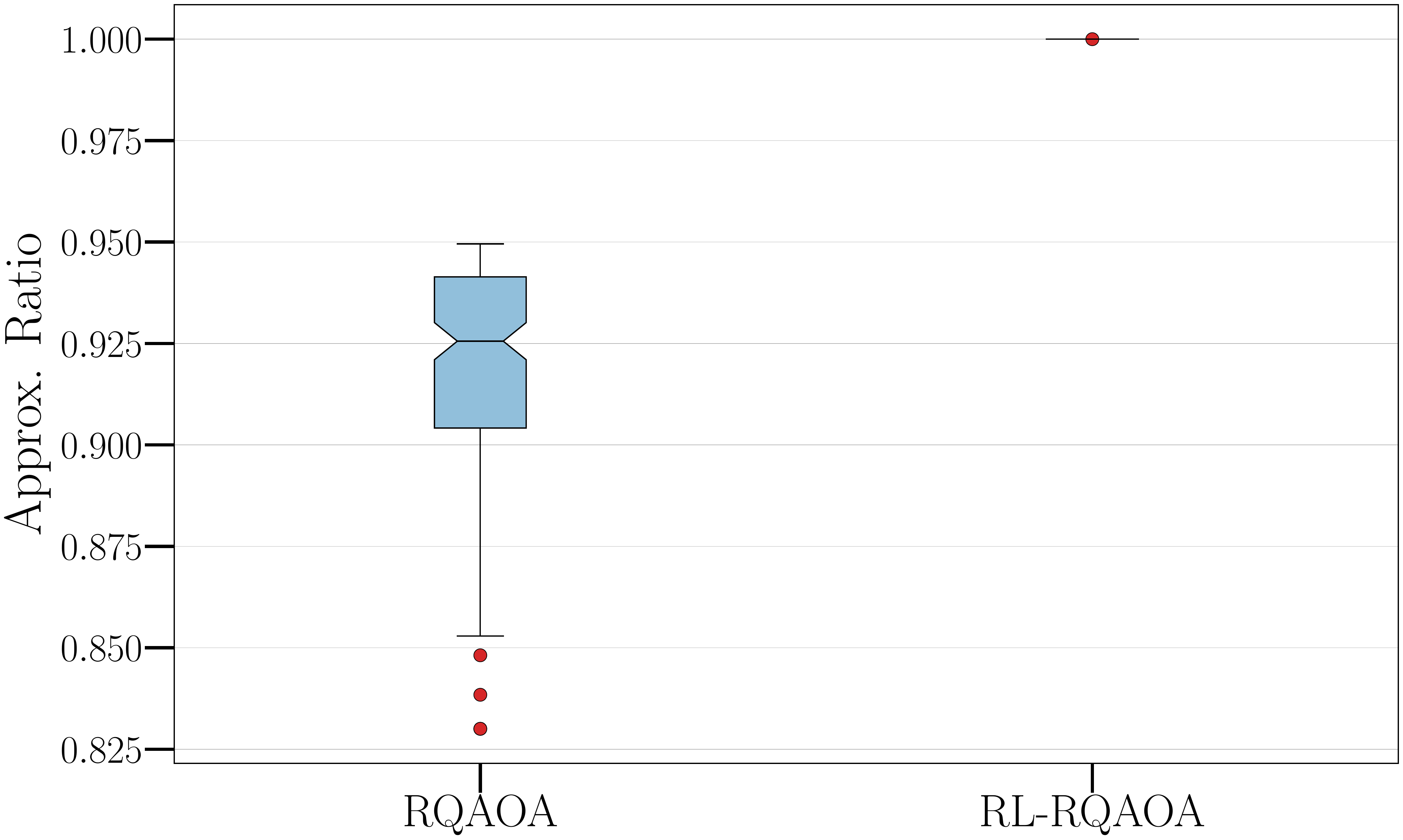}
    \caption{\textbf{Numerical evidence of the advantage of RL-RQAOA over RQAOA in terms of approximation ratio on hard instances}. The box plot is generated by taking the mean of the best approximation ratio over 15 independent runs of 1400 episodes for RL-RQAOA. The RL-RQAOA clearly outperforms RQAOA in terms of approximation ratio for the instances considered (these are exactly the instances where RQAOA's approx. ratio \(\leq 0.95)\). We chose \(n_c=8\) in our simulations and the parameters \(\theta = (\alpha, \gamma, \vec{\beta})\) of the RL-RQAOA policy were initialized by setting \(\vec{\beta} = \{25\}^{{(n^2-n)}/2}\) and the angles \(\{\alpha, \gamma\}\) (at every iteration) were initialized randomly. All agents were trained using \(\mathsf{REINFORCE}\) (Alg. \ref{alg:reinforce}).}
    \label{fig:rl_rqaoa_vs_rqaoa_hard}
\end{figure}

\begin{figure*}
    \centering
    \includegraphics[width=2\columnwidth]{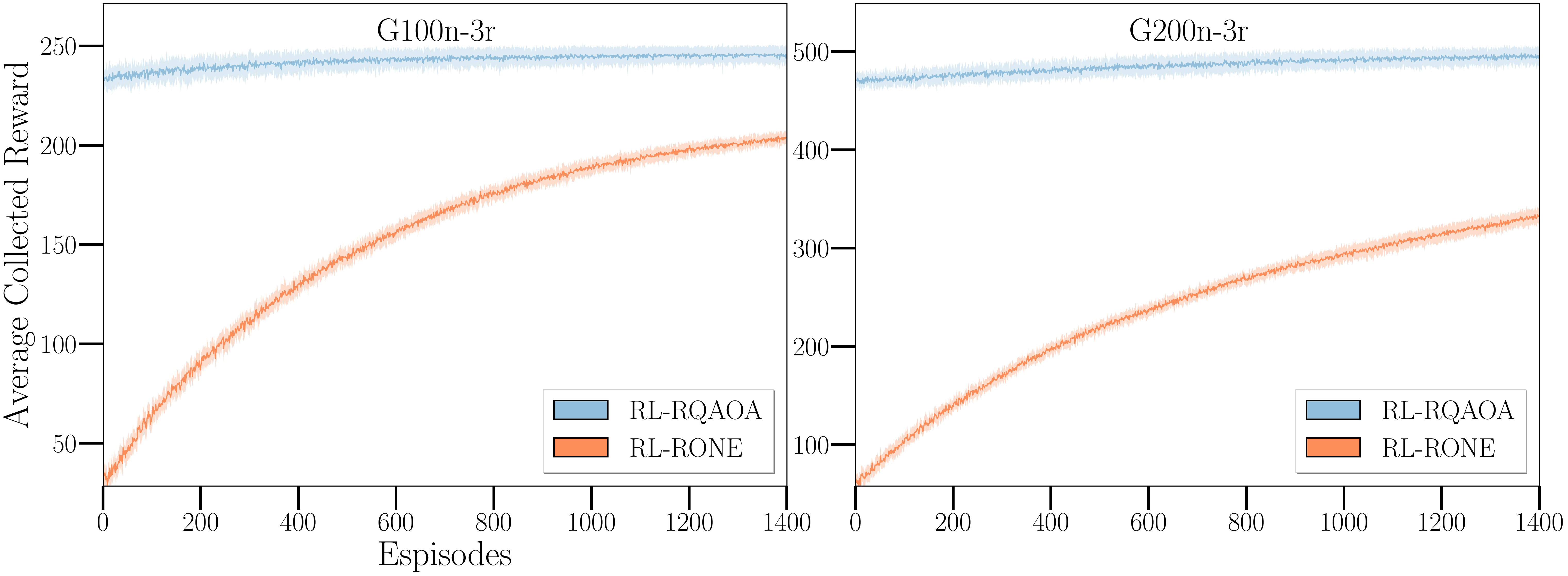}
    \caption{\textbf{Numerical evidence of relevance of the quantum circuit in RL-RQAOA and separation between RL-RQAOA and RL-RONE.} The above plot illustrates the separation between learning curves of RL-RQAOA and RL-RONE agents averaged across 15 bimodal weighted random 3-regular graphs with 100 (left) and 200 (right) nodes each. We chose \(n_c=10\) and \(n_c=18\) for 100 and 200 nodes, respectively in our simulations and the parameters \(\theta = (\alpha, \gamma, \vec{\beta})\) of the RL-RQAOA policy were initialized by setting \(\vec{\beta} = \{25\}^{{(n^2-n)}/2}\) and the angles \(\{\alpha, \gamma\}\) (at every iteration) to energy-optimal angles (i.e., by following one run of RQAOA). All agents were trained using \(\mathsf{REINFORCE}\) (Alg. \ref{alg:reinforce}).}
    \label{fig:rl_rqaoa_vs_rl_rone}
\end{figure*}

\subsection{Benchmarking}\label{subsec:benchmarking}

\subsubsection{RQAOA vs RL-RQAOA on Cage Graphs}
In our first set of experiments, illustrated in Fig.~\ref{fig:rl_rqaoa_vs_rqaoa_random}, we compare the performance of RL-RQAOA with RQAOA on random Ising instances derived from \((d,g)\)-cage graphs \((3 \leq d \leq 7; 5 \leq g \leq 12;~\text{edge weights}~\{-1, +1\})\). 
The aim of this experiment is twofold: first, to show that RL-RQAOA does not perform much worse than RQAOA on instances where the latter performs quite well;
second, to test the advantage of RL-RQAOA over RQAOA in terms of the probability of attaining the optimal solution when there are many ties between two-correlation coefficients \(M_{u,v}\) at every iteration. 
Notably, we already demonstrated earlier (see Fig.~\ref{fig:cage_ties}) that for cage graphs, RQAOA has a constant number of ties between maximal two-correlation coefficients for the majority of the \(n-n_c\) iterations. 
For assessing our hypotheses, we evaluate the average learning performance over \(15\) independent RL-RQAOA runs over 1400 episodes.
In order to fairly compare RL-RQAOA with RQAOA, we run RQAOA independently for 1400 runs and choose the best solution from the result these runs.
Note that, this is a more powerful heuristic than the vanilla-RQAOA (which outputs the first solution it finds) where the hyperparameter (the number of independent runs) controls the solution quality. 
Both RL-RQAOA (vote variant) and RQAOA fail to reach the optimum for \((3, 12)\)-cage graph within the given budget (see Fig.~\ref{fig:rl_rqaoa_vs_rqaoa_random}).
However, by evaluating the resulting learning curves of RL-RQAOA, both our hypotheses can be confirmed for most of the instances.

\subsubsection{RQAOA vs RL-RQAOA on hard instances}
 For the next set of experiments, presented in Fig.~\ref{fig:rl_rqaoa_vs_rqaoa_hard}, the flavour here is similar to the previous experiment but with the aim to show separation between RL-RQAOA and RQAOA for \emph{hard} instances found in Sec.~\ref{subsec:hard_inst}. 
 More specifically, we show that RL-RQAOA always performs better than RQAOA on these instances in terms of the best approximation ratio achieved. 
 We do this by evaluating average learning performance over \(15\) independent RL-RQAOA runs to assess this claim. Interestingly, RL-RQAOA outperformed RQAOA even when the angles of the QAOA circuit were initialized randomly.

 \subsubsection{RL-RQAOA vs RL-RONE}
 
 However, the results in the previous two subsections do not indicate the importance of the quantum part in the quantum-classical arrangement. 
 To address this, we performed a third set of experiments, presented in Fig.~\ref{fig:rl_rqaoa_vs_rl_rone} where both algorithms were tested on random 3-regular graphs of 100 and 200 nodes. 
 By comparing the performance of RL-RONE with RL-RQAOA, we can see a clear separation between learning curves of the agents of these algorithms, highlighting the effectiveness of the quantum circuits in solving the Ising problem.

\section{Discussion and Conclusion}\label{sec:disc}

In this work, we analyzed the bottlenecks of a non-local variant of QAOA, namely recursive QAOA (RQAOA), and based on this, propose a novel algorithm that uses reinforcement learning (RL) to enhance the performance of the RQAOA (RL-RQAOA). 
In the process of analyzing the bottlenecks of RQAOA for the Ising problem, we find small-size \(hard\) Ising instances from a graph ensemble of random weighted \(d\)-regular graphs. 
To avoid missing out on better optimal solutions at every iteration, we cast the variable elimination problem within the RQAOA as a reinforcement learning framework; we introduce a quantum (inspired) policy of RL-RQAOA, which controls the task of switching between exploitative or exploratory behaviour of RL-RQAOA. 
We demonstrate via numerical simulations that formulating RQAOA into the RL framework boosts the performance and performs as well as RQAOA on random instances and beats RQAOA on all hard instances we have identified.
Finally, we note that all the numerical simulations for RQAOA (depth-1) and the proposed hybrid algorithm RL-RQAOA (depth-1) were performed classically, and no quantum advantage is to be expected unless we simulate both of them at higher depths.
An interesting follow-up to this work would be to assess the performance of both RQAOA and RL-RQAOA at higher depths on an actual quantum processing unit (QPU) in both noise and noise-free regimes.

\section*{Acknowledgements}

YJP would like to thank Simon Marshall and Charles Moussa for useful discussions.
SJ would like to thank Hans Briegel for useful discussions in the early phases of this project.  
The authors thank Adri{\'a}n P{\'e}rez-Salinas, and Andrea Skolik for useful comments on an earlier version of this manuscript and Casper Gyurik for reading the final version of this manuscript. 
TB, VD, and YJP acknowledge support from TotalEnergies. 
SJ acknowledges support from the Austrian Science Fund (FWF) through the projects DK-ALM:W1259-N27 and SFB BeyondC F7102. SJ also acknowledges the Austrian Academy of Sciences as a recipient of the DOC Fellowship. The computational results presented here have been achieved in part using the LEO HPC infrastructure of the University of Innsbruck and DSlab infrastructure of the Leiden Institute of Advanced Computer Science (LIACS) at Leiden University.
This work was in part supported by the Dutch Research Council (NWO/OCW), as part of the Quantum Software Consortium programme (project number 024.003.037). 
VD acknowledges the support by the project NEASQC funded from the European Union’s Horizon 2020 research and innovation programme (grant agreement No 951821).
VD also acknowledges support through an unrestricted gift from Google Quantum AI.

\section*{Author contributions}
YJP and SJ contributed equally to this work. YJP, SJ, and VD designed all the experiments. The manuscript was written with contributions from all authors. All authors read and approved the final manuscript.

\section*{Availability of data and materials}%% if any
The datasets and the code used and/or analysed during the current study is available at \url{https://github.com/Zakuta/RL-RQAOA-paper-code/}.

% \section*{Competing interests}
% The authors declare that they have no competing interests.

\bibliographystyle{apalike}
\bibliography{biblography}

\onecolumngrid
\clearpage

\appendix

\section{Theorem on Classical Simulability of Ising problem for depth-1 QAOA}\label{app:classical_sim_qaoa}

\begin{theorem}\label{thm:analytical_form}(~\cite{bravyi2020obstacles, ozaeta2021expectation})
Given an Ising cost Hamiltonian \(H_n = \sum_{u \in V} h_u Z_u + \sum_{(u,v) \in E} J_{uv} Z_u Z_v\). Define \(s(x) := sin(x)\) and \(c(x) := cos(x)\). Then for a fixed pair of qubits \(1 \leq u \leq v \leq n\),
\begin{eqnarray}
    \langle Z_u \rangle_1 = s(2 \alpha) ~ s(2 h_{u}) \prod_{k \neq u} c(2 J_{uk}),
    \label{eq:exp_external}
\end{eqnarray}
\begin{eqnarray}
 \langle Z_uZ_v \rangle_1 = H_A + H_B
\end{eqnarray}
where, 
\begin{equation*}
    \begin{split}
        H_A =  (s(4 \alpha)/2) ~ s(2 J_{uv})\Bigl[c(2 h_{u}) \prod_{k \neq u, v} c(2 J_{uk}) +\\
         c (2 h_{v}) \prod_{k \neq u, v} c(2 J_{vk}) \Bigr]
    \end{split}
\end{equation*}
and 
\begin{equation*}
    \begin{split}
            H_B = (s^2(2 \alpha)/2)\Bigl[c(2 (h_{u}+h_{v})) \prod_{k \neq u, v} c(2(J_{u k}+J_{v k})) -  \\
                c(2 (h_{u}-h_{v})) \prod_{k \neq u, v} c(2(J_{u k}-J_{v k}))\Bigr].
    \end{split}
\end{equation*}
Here, w.l.o.g we assume that the underlying graph is a complete graph \(K_n\) and \(\gamma = 1\) since it can be absorbed into the definition of adjacency matrix \(A\) of the graph.
\end{theorem}

\section{Policy Gradient Method}\label{app:pgm}
This appendix provides a more detailed description of the policy gradient method used to find an optimal policy that simultaneously optimizes variational parameters QAOA and selects a decision variable to eliminate within the RL-RQAOA.

The crux of policy gradient methods lies in (i) a parameterized policy \(\pi_{\theta}\), which drives an agent’s action in an environment, and (ii) value function \(V_{\pi_{\theta}}\) that evaluates long-term performance associated with a policy \(\pi_{\theta}\). 
Policy gradient methods employ a simple optimization approach; i.e., they start with an initial policy \(\pi_{\theta}\) and update the parameters of the policy iteratively by using a gradient ascent algorithm such that the value function \(V_{\pi_{\theta}}(s)\) associated with it is maximized.
This approach can be efficiently applied if one can either evaluate the value function of the policy or at least its gradient \(\grad_\theta V_{\pi_{\theta}}\). In the case of policy gradient methods, the gradient of the value function \(\grad_\theta V_{\pi_{\theta}}\) can be evaluated analytically using Monte Carlo rollouts within an environment. We formally state this in Theorem~\ref{thm:pgm}. \hfill \break

\noindent\textbf{Policy gradient Theorem:} In practice, the value function can be estimated via a Monte Carlo approach: (i) collect \(N\) samples of episodes \(\tau\) of interactions governed by policy \(\pi_{\theta}\) within an environment; (ii) compute expected return \(R(\tau)\) of each episode as in (\ref{eqn:value_func}); and (iii) average out the results. Then the Monte Carlo estimate of value function can be written as
\begin{eqnarray}\label{eqn:value_func_est}
\tilde{V}_{\pi_{\theta}}(s) = \frac{1}{N} \sum_{i=1}^{N} \sum_{t=0}^{H-1} \gamma^t r_{i,t}
\end{eqnarray}

\begin{theorem}(Policy Gradient Theorem~\cite{sutton1999policy})\label{thm:pgm}
Given an environment defined by its dynamics \(\mathbb{P}_E\) and a parameterized policy \(\pi_{\theta}\), the gradient of the value function as defined in (\ref{eqn:value_func}), w.r.t. \(\theta\), is given by
\begin{equation}\label{eqn:grad_value_func}
\begin{split}
    \grad_{\theta} V_{\pi_{\theta}}(s_0) &=\mathbb{E}_{\pi_{\theta}, \mathbb{P}_{E}}\left[\sum_{t=0}^{H-1} \grad_{\theta} \log \pi_{\theta}(a_{t}|s_{t}) V_{\pi_{\theta}}(s_{t})\right] \\
&=\mathbb{E}_{\pi_{\theta}, \mathbb{P}_{E}}\left[\sum_{t=0}^{H-1} \frac{\grad_{\theta} \pi_{\theta} (a_{t}|s_{t})}{\pi_{\theta}(a_{t}|s_{t})} V_{\pi_{\theta}}(s_{t})\right]
\end{split}
\end{equation}
\end{theorem}

This theorem enables the estimation of the gradient of the value function analytically, whose evaluation in the context of sample complexity scales only logarithmic in the parameters \(\theta\) of the policy~\cite{kakade2003sample}.

Moreover, one can estimate the value function terms \(V_{\pi_{\theta}}(s_t)\) in (\ref{eqn:grad_value_func}) by collecting rewards from the Monte Carlo rollouts (as defined in (\ref{eqn:value_func_est})). 
This learning algorithm is called the Monte Carlo Policy Gradient algorithm, otherwise known as \(\mathsf{REINFORCE}\)~\cite{sutton2018reinforcement, williams1992simple}. 
In the literature, there exist other sophisticated approaches, such as the actor-critic method~\cite{konda1999actor}, where the value function is estimated using an additional approximator such as a deep neural network (DNN).

\section{Implementation Details of Algorithms}\label{app:implementation}

This appendix provides specifications for simulating RQAOA, RL-RQAOA, RL-RONE, and Gurobi optimizer. \hfill \break

\noindent\textbf{RQAOA:} The authors in~\cite{bittel2021training} prove that finding optimal parameters in QAOA is an \(\mathsf{NP}\)-hard problem even with logarithmically many qubits at depth 1.
We also often found that the landscape had many extrema and saddle points detrimental to gradient-based methods in QAOA and RQAOA. Hence, we use a brute force search to optimize the variational parameters to alleviate this problem.
To perform the brute force search for depth-1 QAOA efficiently, we show in Appendix~\ref{app:var_opt} that for any fixed value \(\gamma \in \mathbb{R}\), one can compute \(\alpha \in \mathbb{R}\) maximizing the energy over \((\alpha, \gamma)\) by solving a system of equations as defined in (\ref{eqn:sys_of_eqn_var_opt}).
We thus chose 2000 equidistant grid points \(\gamma_1, \ldots , \gamma_{2000}\) in the interval \([0, 2\pi]\) for \(\gamma\).
After finding the grid point \(\gamma_k\) that maximizes the energy, we performed another refined local optimization with an off-the-shelf optimizer COBYLA in the interval \([\gamma_{k-1}, \gamma_{k+1}]\). 
Finally, we note that the optimal angles of QAOA for random graphs in \(\mathcal{G}[n, d, w]\) concentrate, which is in line with several theoretical as well as empirical results in the literature~\cite{brandao2018fixed, chou2021limitations, lotshaw2021empirical, wurtz2021fixed, shaydulin2022parameter, chmoussa2022tranfer}. Throughout, we choose \(n_c = 8\) otherwise specified explicitly. \hfill \break

\noindent\textbf{RL-RQAOA:} Here, we discuss some design choices for RL-RQAOA as it is not entirely clear which one of them has a positive influence on the learning performance of the policy of RL-RQAOA. Firstly, there can be three choices to define \(\vec{\beta}\) that all recover the RQAOA policy for large \(\vec{\beta}\): 
\begin{itemize}
    \item[(i)] \(\mathsf{(\beta-all)}\) \(\vec{\beta} = \{\beta\}^{(n-n_c)}\), i.e., only one parameter among all edges for every iteration of RL-RQAOA;
    \item[(ii)] \(\mathsf{(\beta-one-all)}\) \(\vec{\beta} = \{\beta_{u,v}\}^{(n^2-n)/2}, ~\forall~ (u,v) \in E\), i.e., for each edge the RL agent learns the value of \(\beta_{u,v}\) accounting to a total of \(\binom{n}{2}~ \beta\)'s; and
    \item[(iii)] \(\mathsf{(\beta-all-all)}\) \(\vec{\beta} = \{\beta_{u,v}\}^{(n-n_c)(n^2-n)/2}\), i.e., similar to (ii) but for every iteration of RL-RQAOA, the \(\beta_{u,v}\) are learnt separately.
\end{itemize}

Secondly, one can initialize the variational angles \(\{\alpha, \gamma\}\) randomly, with extremum points or with optimal QAOA angles at every iteration, and then train an agent to learn the angles. 
However, in our simulations, we always warm start the RL agent with optimal QAOA angles to better capture the power of the quantum circuits. 
We call this model a \(\mathsf{WS-RL-RQAOA_{\alpha, \gamma, \beta}}\) where the subscript highlights the parameters for which an agent is trained. 
Following the above nomenclature, \(\mathsf{WS-RL-RQAOA_{\alpha, \gamma, \beta}}\) and \(\mathsf{WS-RL-RQAOA_{\beta}}\) are different models of RL-RQAOA where both of them have been warm started using QAOA angles but only differ in their learning procedures; i.e., an agent for the former trains for optimizing both the variational angles and \(\vec{\beta}\), and the latter only optimizes \(\vec{\beta}\) with fixed optimal QAOA angles.

Empirically, we found that an agent with the configuration of \(\mathsf{WS-RL-RQAOA_{\alpha, \gamma, \beta}}\) with \(\mathsf{\beta-one-all}\) configuration is enough to beat RQAOA (in fact, often solve optimally) within 1400 episodes (pre-defined) at least for all the graphs with less than 30 vertices. 
Hence, we use the choice of \(\vec{\beta}\) as \(\mathsf{(\beta-one-all)}\) in the rest of the manuscript unless specified explicitly. 

Finally, we mention the hyper-parameters with which we performed our simulations. 
We train an agent to choose both the set of variational angles and inverse-temperature constants using the policy gradient method. We set the discounted factor \(\gamma = 0.99\), \(\mathsf{(\beta-one-all)}\) \(\vec{\beta} = \{25\}^{(n^2-n)/2}\) and use ADAM~\cite{kingma2014adam} as an optimizer with learning rates \(\{{lr}_{angles}, {lr}_{betas} = 0.001, 0.5\}\). 
During our hyper-parameter sweep, we noticed that higher values of trainable parameters \(\vec{\beta}\) hampered the learning performance of the RL agents, and this is likely due to their inability to explore the environment. This suggests that for \(\vec{\beta} \rightarrow \infty\), RL-RQAOA mimics the behaviour of RQAOA. In other words, RL-RQAOA is indeed a generalized variant of RQAOA. \hfill \break

\noindent\textbf{RL-RONE:} We simulated RL-RONE with the same set of hyperparameters and configuration as in RL-RQAOA. The only difference is that there are no angles to be learned as the two correlations \(M_{u,v} = 1\) for all \(\{u,v\} \in E\) by design. \hfill \break

\noindent\textbf{EXACT:} The exact solutions were computed using the state-of-the-art commercial solver \emph{Gurobi 9.0} with variable \(MIPGap = 0\) because otherwise the optimization would end prematurely and return a sub-optimal solution.

\section{Variational Optimization of depth-1 QAOA}\label{app:var_opt}

Given a graph $G_n = (V,E)$ with $|V| = n$ vertices,
\begin{eqnarray}
    H_b = \sum_{u \in V} X_u ~ \text{and} ~ H_n = \sum_{(u,v) \in E} J_{u,v} Z_u Z_v.
\end{eqnarray}
Now, we want to compute maximum expected energy \(\langle H_n \rangle_1 := \bra{\Psi_1({\alpha}, {\gamma})} H_n \ket{\Psi_1({\alpha}, {\gamma})}\) at optimum values of $(\alpha, \gamma)$.

Consider an edge $(u,v) \in E$. We only focus on the contribution of mixer Hamiltonian $H_b$, that is, using Pauli operator commutation rules, we expand the inner conjugation \(U_m^\dagger(\alpha)Z_uZ_vU_m(\alpha) = \exp\left(2i\alpha(X_u + X_v)\right) Z_uZ_v\). Then using \(X^2 = I\) this becomes, 
\begin{equation}
    \begin{split}
        [\cos(2\alpha)Z_u + \sin(2\alpha)Y_u] [\cos(2\alpha)Z_v + \sin(2\alpha)Y_v] \\
        = \cos^2(2\alpha)Z_uZ_v + \cos(2\alpha)\sin(2\alpha)[Z_uY_v + Y_uZ_v] + \\ \sin^2(2\alpha)Y_uY_v \\
        = \frac{1+\cos(4\alpha)}{2}Z_uZ_v + \frac{\sin(4\alpha)}{2}[Z_uY_v + Y_uZ_v] +\\ \frac{1-\cos(4\alpha)}{2}Y_uY_v
    \end{split}
\end{equation}

By linearity of expectation, we can write the expectation as,
\begin{eqnarray}
      \langle H_n \rangle_1 = p \cos(4 \alpha)+ q\sin(4 \alpha) + r
\end{eqnarray}
where \(p, q, r\) are real coefficients which are unknown complicated functions of $\gamma$. We can compute these aforementioned coefficients from the following system of equations.
\begin{equation}
    \begin{split}
       \langle H_n \rangle_{1, \alpha = \frac{\pi}{8}, \gamma=\gamma} &=q+r \\
        \langle H_n \rangle_{1, \alpha = \frac{-\pi}{8}, \gamma=\gamma} &=-q+r \\
        \langle H_n \rangle_{1, \alpha = 0, \gamma=\gamma} &=p+r
    \end{split}
\end{equation}
After the values of \(p,q,r\) are known, we can then compute \(\langle H_n \rangle_{1}\) over all $\alpha$ by employing elementary trigonometry,
\begin{eqnarray}
      \max _{\alpha} \langle H_n \rangle_{1} =r+\sqrt{p^{2}+q^{2}}
\end{eqnarray}
where optimal \(\alpha\) can be computed by solving 
\begin{equation}\label{eqn:sys_of_eqn_var_opt}
    \begin{split}
        \tan \left(4 \alpha\right) &=q / p \\
p \cos \left(4 \alpha\right) & \geq 0 \\
q \sin \left(4 \alpha\right) & \geq 0
    \end{split}
\end{equation}

\end{document}